\documentstyle[12pt,twoside]{article}
{\catcode `\@=11 \global\let\AddToReset=\@addtoreset}
\AddToReset{equation}{section}

\def\greaterthansquiggle{\raise.3ex\hbox{$>$\kern-.75em\lower1ex\hbox{$\sim$}}}
\def\lessthansquiggle{\raise.3ex\hbox{$<$\kern-.75em\lower1ex\hbox{$\sim$}}}
\newcommand{\beq}{\begin{equation}}
\newcommand{\eeq}{\end{equation}}
\newcommand{\beqa}{\begin{eqnarray}}
\newcommand{\eeqa}{\end{eqnarray}}
\newcommand{\beqan}{\begin{eqnarray*}}
\newcommand{\eeqan}{\end{eqnarray*}}
\newcommand{\ba}{\begin{array}}
\newcommand{\ea}{\end{array}}
\newcommand{\no}{\nonumber}

\newcommand{\wt}{\widetilde}

\newcommand{\A}{{\cal A}}

\newcommand{\D}{{\cal D}}

\newcommand{\F}{{\cal F}}
\newcommand{\G}{{\cal G}}
\newcommand{\Ha}{{\cal H}}

\newcommand{\M}{{\cal M}}

\def\nz{\ifmmode {I\hskip -3pt N} \else {\hbox {$I\hskip -3pt N$}}\fi}
\def\zz{\ifmmode {Z\hskip -4.8pt Z} \else
       {\hbox {$Z\hskip -4.8pt Z$}}\fi}
\def\qz{\ifmmode {Q\hskip -5.0pt\vrule height6.0pt depth 0pt
       \hskip 6pt} \else {\hbox
       {$Q\hskip -5.0pt\vrule height6.0pt depth 0pt\hskip 6pt$}}\fi}
\def\rz{\ifmmode {I\hskip -3pt R} \else {\hbox {$I\hskip -3pt R$}}\fi}
\def\cz{\ifmmode {C\hskip -4.8pt\vrule height5.8pt\hskip 6.3pt} \else
       {\hbox {$C\hskip -4.8pt\vrule height5.8pt\hskip 6.3pt$}}\fi}
\def\au{{\setbox0=\hbox{\lower1.36775ex%
\hbox{''}\kern-.05em}\dp0=.36775ex\hskip0pt\box0}}
\def\ao{{}\kern-.10em\hbox{``}}

\normalsize
\begin{document}
\bibliographystyle{plain}

\begin{titlepage}
\begin{flushright}
UWThPh-2003-7
\end{flushright}
\vspace{2cm}
\begin{center}
{\Large \bf QED Revisited: Proving Equivalence Between Path Integral 
and Stochastic Quantization} \\[40pt]
Helmuth H\"uffel* and Gerald Kelnhofer\\
Institut f\"ur Theoretische Physik \\
Universit\"at Wien \\
Boltzmanngasse 5, A-1090 Vienna, Austria
\vfill

{\bf Abstract}
\end{center}

We perform the stochastic quantization of scalar  QED based on a generalization of the stochastic gauge 
fixing scheme and its geometric interpretation. It is shown that the stochastic quantization scheme
exactly agrees with the usual path 
integral formulation.

\vfill
\begin{enumerate}
\item[*)] Email: helmuth.hueffel@univie.ac.at
\end{enumerate}
\end{titlepage}

\section{Introduction}
The stochastic 
quantization scheme of Parisi and Wu \cite{Parisi+Wu} has been applied to QED since 
many years. Nice  
agreement with  
conventional calculations was found in several explicit 
examples (for reviews see e.g. 
\cite{Damgaard+Huffel,Namiki}), a general equivalence proof so far was 
lacking.

The main idea of 
``stochastic quantization''  is to 
view Euclidean quantum field theory as the equilibrium limit of 
a statistical system coupled to a thermal reservoir. This system 
evolves in a new additional time direction which is called 
stochastic time until it reaches the equilibrium limit for infinite 
stochastic time. In the equilibrium limit the stochastic 
averages become identical to ordinary Euclidean vacuum 
expectation values.

There are two equivalent formulations of stochastic quantization: In one 
formulation all fields have an additional dependence on stochastic 
time. Their stochastic time evolution is determined by a Langevin 
equation which has a drift term constructed from the gradient of 
the classical action of the system. The expectation values of 
observables are obtained by ensemble averages over the Wiener 
measure.

Equivalently one has a Fokker Planck equation for 
the probability distribution characterizing the stochastic evolution 
of the system. Now expectation values of  observables are 
defined in terms of functional integrals  with respect to the stochastic 
time dependent Fokker-Planck probability distribution. The   
equilibrium
limit of the  probability distribution provides the Euclidean path integral density.

One of the most interesting aspects of this new quantization 
scheme lies in 
its rather unconventional treatment of gauge field theories, in specific of
Yang-Mills theories. We   recall that 
originally it was 
formulated  by Parisi and Wu \cite{Parisi+Wu} without 
the introduction of gauge fixing terms and without the usual 
Faddeev-Popov ghost 
fields; later on a modified approach named stochastic gauge fixing
was given by Zwanziger \cite{Zwanziger81}; further generalizations 
and a globally valid path integral were advocated in \cite{Physlett,annals,annalsym,global}.

The main  difficulty for providing  an equivalence proof in the case of QED
appears to be  a rather nontrivial topological obstruction; all 
previous attempts   failed in 
the past years to 
 identify  the standard - gauge 
fixed- QED action  as  a  Fokker--Planck equilibrium 
distribution.

In this paper we  introduce 
new modifications  of the original Parisi--Wu stochastic process of 
QED, 
yet keeping  expectation values 
of gauge invariant observables  unchanged: The modified 
stochastic process not only has damped flows along the
vertical direction \cite{Zwanziger81} but also is modeled on a specific
manifold  of gauge and matter fields
with associated  {\it flat} connection \cite{Physlett,annals,annalsym,global}. 
It is precisely in this case that the
standard - gauge  fixed - QED path integral density
 can indeed be identified with the weak equilibrium limit of the underlying 
Fokker--Planck probability distribution.

In section 2 the geometrical setting for  QED is 
introduced and the associated 
bundle structure of the space of gauge potentials and matter fields is 
summarized. We 
introduce adapted coordinates, the corresponding vielbeins and metrics.

The generalized stochastic 
process for  QED is presented in section 3, 
 section 4 is devoted to the derivation of the conventional  QED
path integral density as the
equilibrium solution. 
\section{The geometrical setting of  QED}
In this section we present the major geometrical structures of  
QED. We collect in a somewhat formal style all the
necessary ingredients which are needed  for a compact and 
transparent formulation of the stochastic quantization scheme of
QED.
\subsection{Gauge Fields} 
Let $P \rightarrow M$ be a principal $U(1)$-bundle over the n-dimensional  
boundaryless  connected, simply connected and compact Euclidean manifold $M$. 
The photon fields 
are regarded as elements of the affine space $\A$ of all 
connections on $P$.
The gauge 
group $\G$ is given by $\G=C^{\infty}(M;U(1))$ with Lie algebra 
 $Lie \,\G=C^{\infty}(M;u(1))$;
 here  $Lie\,(U(1))=u(1)=i{\bf R}$. The  action of $\G$ on $\A$ is 
defined by
\beq
A\rightarrow 
A^g=A+ g^{-1}dg.
\eeq
Let us define the subgroup $\G_{0} \subset \G$ where 
$\G_{0}=\G/U(1)$ denotes the group of all gauge transformations reduced 
by the constant ones. Since $\G_{0}$ acts freely on $\A$ we consider the principal $\G_{0}$ bundle
$\hat \pi:\A\rightarrow \M=\A/\G$.
One can prove that $\A \rightarrow \M$ is trivializable,  a global section $\hat\sigma: \M\rightarrow \A$ 
being given  
by $\hat\sigma([A])=A^{\omega(A)^{-1}}$.
Here $\omega(A) \in \G_{0}$ is defined by 
\beq
\label{omega}
\omega(A)=exp[ \triangle^{-1} 
d^{*}(A-A_{0})]
\eeq
where $\triangle$ denotes the invertible Laplacian which is acting on the 
Lie algebra  $Lie\G_{0}$\,; $A_{0} \in \A$ is a chosen fixed background 
connection. Note that 
$\omega(A)$  fulfills  
\beq
\label{omegatrafo}
\omega(A^g)=\omega(A)\,g
\eeq
with $g\in \G_{0}$.

\subsection{Matter Fields}

In order to discuss scalar matter fields $\phi$ we chose a representation 
$\rho$ of $g \in \G_{0}$ on the vector
space $V={\bf C}$ ; $\rho(g)$ simply denotes  
multiplication with $g$. We consider the associated 
vector bundles $E=P\times_{\rho}V$ on $M$. Scalar fields are 
described by 
appropriately chosen sections  of $E$.
 In the following we denote by $\F$ the space of 
scalar fields.
 
The action of $\G_{0}$ on ${\Phi}^i:=(A,\phi) \in \A\times \F$ is given by
\beq
\label{trafo}
{\Phi}^i=(A,\phi) \longrightarrow 
({\Phi}^{g})^i=(A^g,\phi^g)=(A+ g^{-1}dg,g^{-1}\phi).
\eeq
We remind that $\A\times\F\stackrel{\pi}{\longrightarrow}\A\times_{\G_{0}}\F$ is 
trivializable iff $\A \stackrel{\hat\pi}{\longrightarrow} \M$ is trivializable. 
Indeed, using the previous construction  of $\omega(A)$
  we obtain a global 
section $\sigma: \A\times_{\G_{0}}\F\rightarrow \A\times\F$ by defining
\beq
\sigma([A,\phi])=(\hat\sigma([A]),\phi^{\omega(A)^{-1}})=(A^{\omega(A)^{-1}},\phi^{\omega(A)^{-1}}).
\eeq
The tangent space  of the configuration space $\A\times\F$ is 
given by $T_{(A,\,\phi)}(\A\times \F)=
\Omega^{1}(M;{  i\bf{\,R}})\times \F$. 
Noting that $v_{\phi} \in 
\F$ by construction transforms equivariantly
we obtain a $\G$-invariant  Riemannian
metric  \,\mbox{$h: T(\A\times\F)\times T(\A\times\F)\rightarrow {\bf 
R}$} by defining
\beq
h_{(A,\phi)}\left( 
(\tau_{A}^1,v_{\phi}^1),(\tau_{A}^2,v_{\phi}^2)\right)=\langle 
\tau_{A}^1,\tau_{A}^2\rangle+\langle v_{\phi}^1,v_{\phi}^2\rangle.
\eeq
Here
\beq
\langle \alpha,\beta\rangle = \frac{1}{2}\int_M (\bar\alpha\wedge *\beta+\alpha \wedge 
*\bar\beta) 
\eeq
and $\alpha, \beta 
\in T\A$, or $T\F$, respectively ; $*$ is the hodge operator 
on $M$,  ${\bar\alpha}$ denotes complex conjugation of $\alpha$.

\subsection{Adapted Coordinates}

Let the globally defined gauge fixing surface $\Sigma$
in $\A\times\F$  be defined by 
\beq
\Sigma=im\,\sigma=\lbrace 
(B,\psi) \in \A\times\F\vert (B,\psi)=(A^{\omega(A)^{-1}},\phi^{\omega(A)^{-1}})\rbrace
\eeq
where $\omega$ is  given by (\ref{omega}). Note that $B$ and $\psi$ 
are invariant under the action of $\G_{0}$ which trivially follows 
from (\ref{trafo}) and from (\ref{omegatrafo}); B satisfies
the ``gauge fixing condition''
\beq
d^{*}(B-A_{0})=0.
\eeq
We define the adapted coordinates 
${\Psi}^{\mu}=\lbrace B,\,\psi,\,g \rbrace$ via the 
bundle maps $\chi :\Sigma \times \G_{0} \rightarrow \A\times\F$ and 
$\chi^{-1} :\A\times\F\rightarrow \Sigma \times \G_{0}$, where
\beq
\chi(B,\psi,g)=(B^g,\psi^g)
\eeq
and 
\beq
\chi^{-1}(A,\phi)=(\sigma([A,\phi]),\omega(A))
=(A^{\omega(A)^{-1}},\phi^{\omega(A)^{-1}}, \omega(A)).
\eeq
The differentials $T\chi$ and $T \chi^{-1}$
 are  calculated straightforwardly (compare also with \cite{annalsym})
\beq
T\chi(\zeta_B,v_{\psi},Y_g) = (\zeta_B + d \theta_g(Y_g),g^{-1} 
(v_{\psi}-\theta_g(Y_g) \Phi)),
\eeq
as well as
\beq
T\chi^{-1}(\tau_A,v_{\phi}) = ({\bf P}\,\tau_A, \omega(A) (v_{\phi}+(\triangle^{-1} 
d^* \tau_A)\phi), \omega(A) \triangle^{-1} 
d^* \tau_A). 
\eeq
Here  $(\zeta_{B}, v_{\psi})  \in T_B \Sigma$, $Y_g \in T_{g} \G$ and  
$(\tau_A,v_{\phi}) \in 
T_{(A,\,\phi)}(\A\times \F)$. The  Maurer-Cartan form 
on $\G_{0}$ is denoted by $\theta$, ${\bf P}$ is the transversal projector  ${\bf P} = {\bf 1} - 
d \triangle^{-1} d^*$. 
From the differentials $T\chi$ and $T \chi^{-1}$ we read off the vielbeins 
$e^i{}_\mu= \frac{\delta {\Phi}^i}{\delta {\Psi}^\mu}$ and their 
 inverses $
E^\mu{}_i = \frac{\delta {\Psi}^\mu}{\delta {\Phi}^i}
$
 corresponding to the change of variables
${\Psi}^{\mu}=\lbrace B,\psi,g\rbrace \leftrightarrow {\Phi}^{i}=\lbrace A,\phi \rbrace$.

Above  we defined a Riemannian structure on $\A\times \F$ in terms of the 
$\G_{0}$ invariant metric $h$. In   adapted coordinates
 this metric  is given now as  the pullback  
$G=\chi^* h$; explicitly we obtain 
\beqa
\lefteqn{G_{(B,\psi,g)} 
((\zeta_B^1,v_{\psi}^1,Y_g^1),(\zeta_B^2,v_{\psi}^2,Y_g^2))
} \no  \\
&=& \langle \zeta_B^1,{\bf P}\zeta_B^2\rangle +
\langle \theta_g(Y_g^1),(\triangle+{\vert \psi\vert}^2) \theta_g(Y_g^2)\rangle +
\langle v_{\psi}^1,v_{\psi}^2 \rangle -
\langle \theta_g(Y_g^1) \psi, v_{\psi}^2\rangle- 
\langle v_{\psi}^1,\theta_g(Y_g^2)\psi\rangle\no.
\\
\eeqa
Here $(\zeta_B^1,v_{\psi}^1), (\zeta_B^2,v_{\psi}^2) \in T_{(B,\psi)} \Sigma$ and $Y_g^1,Y_g^2 \in T_g \G$.
In matrix notation we have $G = e^{\dagger}\,e$ and $
G^{-1}= E\,
 E^{\dagger}$.
The determinant of $G$ is  given by $det \,G= \triangle$.

\section{Parisi-Wu Stochastic Quantization}
For scalar QED we have 
\beq
S_{inv}=\langle D_{A} \varphi,D_{A} \varphi \rangle + 
m^2\langle  \varphi, \varphi \rangle +
\frac{1}{2} \langle F,F \rangle
\eeq
where $D_{A} \varphi=(d-A)\varphi$ and $F=dA$. The Parisi-Wu Langevin 
equations are given by
\beq
d\A=-\frac{\delta S_{inv}}{\delta \A} ds+dU
\eeq
\beq
d\varphi=-\frac{\delta S_{inv}}{\delta \bar \varphi} ds+dV
\eeq
where the Wiener increments fulfill
\beq
dU\,dU=2 ds,\quad dV\, d\bar V=2 ds.
\eeq
This can be summarized by 
\beq
d{\Phi} = - \frac{\delta S_{inv}}{\delta {\Phi}}
 ds + d \xi,
\quad d\xi\,d\xi^{\dagger}=2 \cdot {\bf 1}\,ds.
\eeq
Using Ito calculus 
\cite{Arnold,belo} we transform the   Parisi-Wu Langevin equations 
into  adapted coordinates 
\beq
d{\Psi} =
  \left[ -  G^{-1}
\frac{\delta S_{inv}}{\delta {\Psi}} 
+ \frac{\delta G^{-1}}{\delta {\Psi}}
  \right] ds
  + d \zeta,
\eeq
where
\beq
d\zeta d \zeta^{\dagger}= 2\,{ G}^{-1}ds.
\eeq
The use of adapted coordinates allows to disentangle the complicated dynamics of gauge 
independent and gauge dependent degrees of freedom; it will be of 
great value later on.
For completeness we note the Fokker-Planck operator in the original 
variables
\beq
L[\Phi] = \frac{\delta}{\delta \Phi}
\left[
\frac{\delta S_{inv}}{\delta \Phi} + \frac{\delta}{\delta \Phi}
\right],
\eeq
as well as in the adapted coordinates
\beq
L[\Psi] =  \frac{\delta}{\delta \Psi}\;
G^{-1} \left[ \frac{\delta S_{inv}}{\delta \Psi} 
+ \frac{\delta}{\delta \Psi}\right].
\eeq
We remark that in the case of the Parisi-Wu  processes 
diffusion along the vertical direction takes place and no equilibrium distribution is approached.
Thus a Fokker-Planck formulation  of the Parisi-Wu
stochastic quantization scheme is impossible: The gauge invariance of the action 
$S_{inv}$ is leading to divergencies along the vertical directions when 
trying to  
normalize the Fokker Planck density.
\section{Generalized Stochastic Quantization}

\subsection{Geometric Obstruction}
Our 
 equivalence proof relies on   specific allowed modifications
of the metric on the field  space, which governs the stochastic process. These modifications correspondingly  are implying changes 
of the  associated 
Fokker-Planck operator. We are going to show  that this can be achieved in such a way 
that the  resulting Fokker--Planck 
operator has a positive kernel and is
 annihilated 
on its  {\it right}  by the standard gauge fixed QED path integral 
density.
In order for this to be the case, however,  a certain 
integrability condition for the drift term of the considered 
stochastic process has to be fulfilled. 
Surprisingly similar as in  the pure Yang--Mills theory 
also in the abelian QED case there appears a violation of this 
condition; it is only after a nontrivial modification of the 
underlying stochastic processes (see next subsection) that this obstruction can be overcome. 

Proceeding step by step we first note (see Zwanziger 
\cite{Zwanziger81}) that a damping force along the gauge 
orbit has to be introduced in order to maintain the probabilistic 
interpretation of the Fokker--Planck formulation. Although  knowing  that 
this additional force will not alter 
expectation values of gauge invariant quantities it is disappointing
to observe that due to its presence the standard - gauge fixed - QED action 
will never   annihilate the 
Fokker--Planck operator on its right side due to the following reason:

We recall that the  bundle metric  
$h_{(A,\phi)}$ on the
associated fiber bundle $\A\times  \F \rightarrow \A\times_\G  \F$ which 
is invariant under the corresponding 
group action gives rise to a natural connection $\gamma$, whose horizontal 
subbundle $\Ha$ is orthogonal to the corresponding group. The horizontal bundle 
	$\Ha[\A\times \F;\gamma]$ with respect to $\gamma$ is defined by \mbox{$\Ha[\A\times 
	\F;\gamma] \perp_{h} V(\A\times\F)$}, where the orthogonality is with 
	respect to 	$h_{(A,\phi)}$. Elements of the vertical bundle 
	$V(\A,\F) \rightarrow  \A\times_\G  \F $ are given 
in the form 
\beq
{Z}_{\xi}(A,\phi)=\frac{d}{dt}\vert_{{}_{{}_{t=0}}}(A^{exp t 
\xi},\phi^{exp t \xi}) =(d \xi,-\xi \phi),
\eeq
where $\xi\in C^{\infty}(M;{i\bf R})$. 
The orthogonal span with respect to the vertical bundle fulfills 
\beq
d^{*}\tau_{A}+\frac{1}{2}( {\bar v}_{\phi}\phi-v_{\phi}\bar\phi)=0, \quad 
\tau_{A}\in T_{A}\A, \, v_{\phi}\in T_{\phi}\F
\eeq
which follows from 
\beq
0=h_{(A,\phi)}\left( 
(\tau_{A},v_{\phi}),{Z}_{\xi}(A,\phi)\right)=h_{(A,\phi)}\left( 
(\tau_{A},v_{\phi}),(d \xi,-\xi \phi)\right)=\langle 
\tau_{A},d \xi\rangle-\langle v_{\phi},\xi \phi\rangle.
\eeq
Explicitly we can prove that $\gamma_{(A,\phi)}(\tau_{A},v_{\phi}) \in 
	C^{\infty}(P;i{\bf R})$
\beq 
\gamma_{(A,\phi)}(\tau_{A},v_{\phi})=(\triangle + |\phi|^2)^{-1} 
[d^{*}\tau_{A}+\frac{1}{2}({\bar v}_{\phi} \phi - v_{\phi} \bar\phi)]
\eeq
defines a  connection induced by 
	$h_{(A,\phi)}$ in the 
principal bundle $\A\times\F \rightarrow  \A\times_\G  \F$ and is $U(1)$ invariant.
 Calculating its curvature $\Omega\left( 
(\tau^1,v^1),(\tau^2,v^2)\right)$
we find that it is
	nonvanishing and given by
	\beqa
\Omega &=&
(\triangle + |\phi|^2)^{-1}(v^2 \bar\phi+{\bar v}^{2}\phi)(\triangle + |\phi|^2)^{-1}
[d^{*}\tau^{1}_{A}+\frac{1}{2}({\bar v}^{1}\phi-v^{1}\bar\phi)]\no\\
&&\mbox{}-(\triangle + |\phi|^2)^{-1}(v^1 \bar\phi+{\bar v}^{1}\phi)(\triangle + |\phi|^2)^{-1}\no\\
&&\mbox{}+[d^{*}\tau^2_{A}
+\frac{1}{2}({\bar v}^{2}\phi-v^{2}\bar\phi)]
(\triangle + |\phi|^2)^{-1}(v^1 {\bar v}^{2}-{\bar v}^{1}v^2)
\eeqa
As a consequence \cite{Physlett,annals,annalsym,global} there does
not exist (even locally) a manifold whose tangent bundle is isomorphic
to this horizontal subbundle. Specifically this  implies that any vector
field along the gauge group cannot be written as a gradient  
with respect to the metric $h_{(A,\phi)}$. 
The total 
drift term - containing the extra vertical force term -
 thus can never arise as  derivative of the standard  gauge fixed  QED action; 
the Fokker-Planck operator can never be annihilated on its right 
by the standard QED path integral density; an equivalence proof 
presently cannot be given.

\subsection{The Induced  Field  Metric with Flat Connection}
The crucial observation  in \cite{Physlett,annals,annalsym,global} is 
to consider a larger class of modified stochastic processes than
considered so far, yet  always keeping expectation values of gauge 
invariant observables unchanged: One introduces not only the extra
vertical   drift terms as discussed above
 but one also  modifies the Wiener increments by specific extra terms and 
 introduces   extra so called 
Ito-terms, correspondingly. 

The idea is to view
the new terms multiplying the Wiener increments
as  vielbeins giving rise to the inverse of a yet
not specified metric  on the space $\A\times \F$.
The appearance of this metric induces a 
specific connection with a potentially  analogous obstruction as  
discussed above. A necessary requirement to 
overcome this 
obstruction is therefore that the corresponding curvature has to vanish.
The question how to find such a metric 
is reduced to the question how to find a flat 
connection.

Indeed,  there exists a flat connection $\wt \gamma$ in our 
bundle.
This connection is the pull-back of the Maurer--Cartan form
$\theta$ on $\G_{0}$ via the global trivialization $\chi^{-1}$ and 
$pr_{\G}$
\beq
\wt \gamma=(\chi^{-1 \,*}pr^{*}_{\G} \theta)(A,\phi)=\triangle^{-1}d^{*},
\eeq
where $pr^{*}_{\G}$ is the projector $\Sigma \times \G \rightarrow 
\G$. The projector onto the horizontal subbundle 
	${\wt\Ha}[\A\times \F;\wt \gamma]$ with respect to $\wt\gamma$ is given
by
\beq
\wt {\bf P} = {\bf 1} - D_A \wt \gamma.
\eeq
We see that the horizontal
subbundle $\wt \Ha$ is orthogonal to the gauge orbits
with respect to the induced field metric; in particular the gauge fixing surface is then
orthogonal to the gauge orbits.

In the adapted coordinates the induced field metric is denoted by $\wt G= 
\wt e^{\dagger}\,\wt e$. The just discussed orthogonality condition of the gauge fixing
surface and the gauge orbit with respect to the  
induced field  metric  is
transformed into simply
\beq
(\wt G^{-1})^{\Sigma \G} = (\wt G^{-1})^{\G \Sigma} = 0, \quad {\rm 
where}\quad  \wt G^{-1}=\wt E\,\wt E^{\dagger}\quad {\rm 
with} \quad  \wt E \,\wt e = {\bf 1}.
\eeq
This condition is fulfilled provided $\wt E$ is defined as
\beq
\wt E=\left( \ba{c}  E^{\Sigma}\\[9pt]
 e_{\G}^{\dagger}\ea \right).
\eeq
To complete our discussion we also have to specify the  vertical drift 
term; it is  related to   the gradient of $S_{\G}$, where we chose
\beq
S_{\G}[g]=\frac{1}{2}\langle d^{*} 
g^{*}\theta^{U(1)},d^{*} g^{*}\theta^{U(1)}\rangle,
\eeq
where $\theta^{U(1)}$ is the Maurer Cartan form on $U(1)$. Note that 
in the original variables we obtain the standard background-gauge gauge fixing term
\beq
(\chi^{-1 \,*}pr^{*}_{\G} S_{\G})(A,\phi)=S_{\G}(\omega(A))=
\frac{1}{2}\langle d^{*} 
(A-A_{0}),d^{*}(A-A_{0})\rangle.
\eeq
Summarizing we   have
\beq
d{\Psi}=
  \left[ - {\wt G}
\frac{\delta S_{tot}}{\delta {\Psi}} 
+ \frac{\delta {\wt G}}{\delta {\Psi}}
 \right] ds 
  + {\wt \zeta}
\eeq
where
\beq
S_{tot}=S_{inv}+ S_{\G}, 
\quad
{\rm and}\quad d{\wt \zeta} d {\wt \zeta}^{\dagger}= 2{\wt G}^{-1}ds.
\eeq

\subsection{The Equivalence Proof}
It is  easy now  to prove for QED the 
equivalence of the stochastic quantization scheme with the path 
integral quantization. For the formulation in terms of the adapted 
coordinates ${\Psi}=\{B,\psi,g\}$ the associated Fokker--Planck equation  is derived in 
straightforward manner
\beq
\frac{\partial \rho[{\Psi},s]}{\partial s} = L[{\Psi}] \,\rho[{\Psi},s],
\eeq
where  the Fokker-Planck operator $L[{\Psi}]$ is 
appearing in just factorized form
\beq
L[{\Psi}] = \frac{\delta}{\delta {\Psi}}
\,\wt G^{-1} \,\left[ \frac{\delta S_{tot}[{\Psi}]}{\delta {\Psi}}
+ \frac{\delta}{\delta {\Psi}}
 \right].
\eeq
Due to the positivity of $\wt G$ the fluctuation dissipation theorem 
applies and the equilibrium Fokker--Planck distribution 
$\rho^{\rm eq}[{\Psi}]$ 
obtains by direct inspection as
\beqa
\rho^{\rm eq}[{\Psi}] &=&
\frac{ e^{- S_{tot}[B,\psi,g]}} {\int_{\bf \Sigma \times \G_{0}} 
\D B \D \psi \D g\,
 e^{- S_{tot}[B,\psi,g]}} \no \\[12pt]
&=& \frac{e^{-S_{inv}[B,\psi]} 
\,e^{-S_{\G}[g]}} {\int_{\bf \Sigma} \D B \D \psi \, e^{-S_{inv}[B,\psi]}
\,\int_{\bf \G_{0}} \D g  \,e^{-S_{\G}[g]}}.
\eeqa
This result is completely equivalent to the standard background-gauge fixed
QED path integral prescription. 
The additional $finite$ contributions of the gauge 
degrees of freedom always cancel out when evaluated on gauge invariant 
observables.

Similarly, in terms of the original variables ${\Phi}=\{A,\phi\}$
 the Fokker-Planck equilibrium distribution 
is given by the standard background-gauge fixed path integral density
\beq
\rho^{\rm eq}[{\Phi}] = \frac{e^{-S_{tot}[{A,\phi}]}}{\int_{\bf 
\A} \D A \D \phi \,e^{-S_{tot}[{A,\phi}]}}
=\frac{e^{-S_{inv}[A,\phi]-S_{\G}(\omega(A))}}{\int_{\bf 
\A} \D A \D \phi \,e^{-S_{inv}[A,\phi]-S_{\G}(\omega(A))}}.
\eeq

\end{document}